\documentstyle[12pt]{article}
\def\bbox{{\,\lower0.9pt\vbox{\hrule \hbox{\vrule height 0.2 cm

\hskip 0.2 cm

\vrule  height 0.2 cm}\hrule}\,}}

\def\bbox{{\,\lower0.9pt\vbox{\hrule \hbox{\vrule height 0.2 cm

\hskip 0.2 cm

\vrule  height 0.2 cm}\hrule}\,}}
\begin{document}
\setlength{\unitlength}{1mm}
\title{{\hfill {\small Alberta-Thy-07-99 } } \vspace*{2cm} \\
Statistical Mechanics on Axially-symmetric Space-times with the
Killing Horizon and Entropy of Rotating Black Holes in Induced Gravity}
\author{\\
V.P. Frolov\thanks{e-mail: frolov@phys.ualberta.ca}
${}^{1}$ and D.V. Fursaev\thanks{e-mail: fursaev@thsun1.jinr.ru}
${}^{2}$
\date{}}
\maketitle
\noindent  
{
 \\ $^{1}${ \em Theoretical Physics Institute,
Department of Physics, \ University of Alberta \\
Edmonton, Canada T6G
2J1 and } \\
$^{2}${\em Joint Institute for Nuclear Research, Bogoliubov
Laboratory of Theoretical Physics, \\
141 980 Dubna, Russia} 
}
\bigskip

\begin{abstract}
We develop a method for computing the free-energy of a canonical
ensemble of quantum fields near the  horizon of a rotating black hole. 
We show that  the density of energy levels of a quantum field on a
stationary  background can be related to the density of levels of the
same field  on a fiducial static space-time. The effect of the
rotation  appears in the additional interaction of the "static" field
with a  fiducial abelian gauge-potential.  The fiducial static
space-time and  the gauge potential are universal, i.e., they are
determined by the  geometry of the given physical space-time and do not
depend on the spin  of the field.  The reduction of the stationary 
axially symmetric problem to the  static one leads to a considerable
simplification in the study of  statistical mechanics and we use it to
draw a number of conclusions.  First, we prove that divergences of the
entropy of scalar and spinor  fields at the horizon in the presence of
rotation have the same form as in the static case and can  be removed
by renormalization of the bare black hole entropy.  Second,  we
demonstrate that statistical-mechanical representation of the 
Bekenstein-Hawking entropy of a black hole in induced gravity is 
universal and does not depend on the rotation.  
\end{abstract}

\bigskip

{\it PACS number(s): 04.60.+n, 12.25.+e, 97.60.Lf, 11.10.Gh}

\baselineskip=.6cm

\noindent
\newpage
\section{Introduction}
\setcounter{equation}0

There are different approaches to the explanation of the
Bekenstein-Hawking entropy $S^{BH}$ of black holes. One of them is to
relate $S^{BH}$ to the  statistical-mechanical entropy $S^{SM}$ of the
thermal atmosphere of quantum fields near the black hole horizon
\cite{ThZu}--\cite{Hooft}. $S^{SM}$ can be naturally interpreted as the
entropy of entanglement \cite{BKLS}--\cite{FrNo:93} which arises as the
result  of quantum correlations on the horizon. For a review and recent
development of these ideas  see \cite{FF:98a} and references therein.

Computation of $S^{SM}$ is a delicate procedure because the density of
energy levels $dn/d\omega$ of single-particle excitations is  divergent
near the horizon. On static space-times  the wave equation for a mode
with the certain energy $\omega$ is similar to a relativistic
Schroedinger equation. In this case, one can define a single-particle
Hamiltonian $H$ as the "square root" of a  Laplace operator acting on a space
whose 3-geometry is approximated by the  hyperbolic geometry. It enables
one to use rigorous methods to  investigate the spectrum $\omega$ of
$H$ and find its density $dn/d\omega$ \cite{CVZ:95}--\cite{F:98}. 
However, this approach cannot be applied directly to fields near a
rotating black hole. The difficulty is that in the latter case the wave
equation which determines the spectrum of energies contains  terms
which are both quadratic and  linear in $\omega$.  As a result, one
cannot define the single-particle Hamiltonian  by taking naively the
"square root" prescription.   

There was no rigorous and universal method how to avoid this difficulty
in computing $S^{SM}$ for rotating black holes. For this reason, many
authors used WKB approximation \cite{LeKi}--\cite{JiYa:a}  or alternative
\cite{Cognola} approaches. These approaches  were always based  on
approximations and, hence, require a justification.  The  calculations
of the entropy were also done by Mann and Solodukhin, who were 
using the Euclidean  formalism \cite{MaSo:96}--\cite{MaSo:97}. In the
recent paper \cite{JiYa:a} Jing and Yan demonstrated an agreement of the
WKB calculations of the black hole entropy for rotating black holes with
the Euclidean results.

\bigskip

The aim of this work is two-fold. First, we suggest a general method
for computing the spectrum and doing statistical-mechanical
computations in the case of rotating black holes. Second, we draw with
its help a  number of consequences concerning the entropy $S^{SM}$. The
proposed method  uses the covariant Schwinger-DeWitt technique and it 
can be applied to fields of  different spins.  

The terms in the wave equation which are
linear in $\omega$ appear due to the non-vanishing component
$g_{t\varphi}$ of  the background metric. Our idea is to include these
terms in the definition of a fiducial single particle Hamiltonian 
$H(\omega)$ which depends on the energy $\omega$ as on additional
parameter. The operator $H(\omega)$ can be interpreted as the 
Hamiltonian of a particle moving on a fiducial background and
interacting with an external fiducial gauge potential  with the only
non-vanishing component  $A_\varphi \sim g_{t\varphi}$.  In some regard
the appearance  of the potential is analogous to the origin of the 
gauge field from the component $g_{5\mu}$ of the metric  in
Kaluza-Klein compactifications.

Thus, our method is to reduce the problem of computations
on the stationary background to computations on a fiducial
static space-time with external gauge field, i.e., to the problem
which is already solved. As we will see, the form of the fiducial
background and gauge field is determined  by the geometry of the
physical space-time only and is the same for the fields of different 
spins.  In this sense, the method is universal. Moreover, the 
method can be applied to black holes in arbitrary dimensions.

\bigskip

The paper is organized as follows. In Section 2 we introduce  the
fiducial background and demonstrate our method for scalar and spinor
fields. In Section 3 we use the obtained results  to derive the
one-loop divergences of the entropy $S^{SM}$ for a rotating black  hole
in a most complete form. We show that this form is the same as in the
case of static space-time and, hence, the known renormalization
procedure \cite{SU}--\cite{DLM:95} can be applied to remove the
divergences. In Section 4 we discuss black hole entropy in models of induced 
gravity \cite{FFZ:97},\cite{FF:98}. We use the obtained results to
prove that the statistical-mechanical form of the Bekenstein-Hawking
entropy $S^{BH}$ for static and rotating black holes in this theory is
the same. This fact may be considered as another evidence that the
mechanism of generating the black hole entropy \cite{FF:97} is
universal. Our concluding remarks are presented  in Section 5. In
Appendix we demonstrate that geometrical  characteristics of the
physical and fiducial backgrounds at the horizon  coincide.  We use
this property in the main text in the paper.

\section{Statistical mechanics in a space-time of a rotating black hole}
\setcounter{equation}0

\subsection{Stationary axisymmetric space-times}

We begin with the formulation of statistical-mechanics of scalar and
spinor fields on a stationary axially symmetric space-times with a
Killing horizon. Consider a $D$-dimensional space-time ${\cal M}$ with
two  commuting  Killing vector fields $\partial_t$ and
$\partial_\varphi$. We assume that  the vector $\partial_t$ is
time-like at asymptotic infinity, and is normalized at infinity by the
condition $\partial_t \cdot \partial_t =-1$. The other Killing vector
$\partial_\varphi$  corresponds to the symmetry of space with respect
to rotation. It commutes with $\partial_t$  and has closed integral
curves. The field $\partial_\varphi$ is nonzero everywhere in the
exterior region and at the horizon, except on the rotation axis. We
also assume that at the rotation axis space-time is locally flat (i.e.,
there is no conical singularities). The vector fields $\partial_t$ and
$\partial_\varphi$ possessing the properties described above are
uniquely defined in a axially symmetric asymptotically flat
space-time. 

In such a space-time one can introduce coordinates $t, \varphi, x^k$
($k=2, \dots ,D-2$) in which the metric takes the form\footnote{
Strictly speaking, this is true for vacuum 4$D$ space-times. In a more
general case, when matter or fields are present, a so-called {\em
circularity condition} must be satisfied. See, e.g. \cite{Kram:80}.
}
\begin{equation}\label{1.6}
ds^2=g_{tt}dt^2+2g_{t\varphi}dt d\varphi +
g_{\varphi\varphi}d\varphi^2
+g_{ik}dx^idx^k~~~.
\end{equation}
Here $0\le \varphi \le 2\pi$, and the components of the metric depend
on coordinates $x^k$ only.

We assume that a stationary asymptotically flat space-time $\cal M$
contains a rotating black hole and is a solution of Einstein equations
with matter satisfying suitable hyperbolic equations. In this case, the
event horizon $H$ coincides with the Killing horizon \cite{HaEl:73}.
The latter is defined as a null surface, $H$, to which a Killing vector
$\xi$ is normal. In the stationary axisymmetric  space-time  the
Killing vector $\xi$ can be written as
\begin{equation}
\label{1.1}
\xi=\partial_t+\Omega_H\partial_\varphi~~~.
\end{equation}
Here $\Omega_H$ is the angular velocity of the black hole which is
constant at the horizon. The position of the horizon $H$ is determined
by the equation
\begin{equation}
\label{1.1a}
(g_{t\varphi})^2-g_{tt}g_{\varphi\varphi}=0~~~,
\end{equation} 
while the angular velocity $\Omega_H$ is
\begin{equation}
\label{1.1b}
\Omega_H=-\left. {g_{t\varphi}\over g_{\varphi\varphi}}\right|_{H}~~~.
\end{equation} 

For our purpose, it is convenient to rewrite metric (\ref{1.6}) in the
coordinates which are rigidly co-rotating with the black hole. Let
\begin{equation}\label{1.10a}
\tilde{\varphi}=\varphi-\Omega_H t~~~,
\end{equation}
then metric (\ref{1.6}) takes the form
\begin{equation}\label{1.10}
ds^2=-N^2dt^2+g_{\varphi\varphi}(d\tilde{\varphi}+
\tilde{N}_\varphi dt)^2
+g_{ik}dx^idx^k~~~.
\end{equation}
Here,
\begin{equation}\label{1.9}
N^2\equiv-{1 \over g^{tt}}={(g_{t\varphi})^2-g_{tt}g_{\varphi\varphi}
\over g_{\varphi\varphi}}~~~,
~~~\tilde{N}^\varphi=N^\varphi+\Omega_H~~~,~~~
N^\varphi\equiv{g_{t\varphi} \over g_{\varphi\varphi}}~~~.
\end{equation}
It is evident that
\begin{equation}\label{1.9.c}
\left. \tilde{N}^\varphi \right|_{H}=0~~~.
\end{equation}
From equation (\ref{1.1a}) it follows that $N^2=0$ on the horizon. (At
the axis of symmetry $N^2$ can be defined by continuity.) By using the
condition of the regularity of the metric on the horizon it is possible
to  show that the ratio $\tilde{N}^\varphi/N^2$  is not  singular on
$H$. 

One can also rewrite the line element (\ref{1.10}) 
in the form which will be especially useful for our purposes
\begin{equation}\label{x1}
ds^2=-B(dt-W d\tilde{\varphi})^2+Cd\tilde{\varphi}^2
+g_{ik}dx^i dx^k=
-B(dt-W d\tilde{\varphi})^2+dl^2
~~~.
\end{equation}
Here
\begin{equation}\label{1.15}
B=-\xi^2=
N^2\left(1-g_{\varphi\varphi}{(\tilde{N}^\varphi)^2 \over N^2}\right)
~~~,
\end{equation}
\begin{equation}\label{1.14}
{1 \over C}={1 \over g_{\varphi\varphi}}\left(1-
g_{\varphi\varphi}
{(\tilde{N}^\varphi)^2\over N^2}\right)~~~,
\end{equation}
\begin{equation}\label{1.16}
W=C{\tilde{N}^\varphi \over N^2}~~~.
\end{equation}

Consider a Killing observer, that is the observer which has the
velocity $u^{\mu}\sim \xi^{\mu}$. Let a point $p$ lying on the
worldline of this observer has coordinates $x^\mu=(t,\tilde{\varphi},
x^i)$. The condition that another event $y^\mu=(t+d t,\tilde{\varphi}+
d\tilde{\varphi}, x^i)$ in its vicinity is simultaneous with $p$, that
is it lies in the plane orthogonal to $\xi$, implies $d t=W
d\tilde{\varphi}$. The spatial distance between these two events is
$dl$.

In the general case, the horizon of a rotating black hole is surrounded
by the region called the ergosphere. Inside the ergosphere $\xi$ is
time-like, while the vector $\partial_t$ is space-like.

\subsection{Scalar fields}

Let us now investigate the properties of the spectrum of
single-particle excitations in a space-time of a stationary rotating
black hole. We begin with the free scalar field which is described by
the Klein-Gordon equation
\begin{equation}\label{1.5}
(-\nabla^\mu\nabla_\mu+V)\phi=0~~~,
\end{equation}
\begin{equation}\label{1.5aa}
V=\xi R +m^2~~~.
\end{equation}
In accordance with the assumed symmetry, we can write a solution of this
equation by using decomposition into modes
\begin{equation} 
\label{1.2} 
\phi_{\omega,l}(t,\varphi,{\bf x})=e^{-i(\omega+\Omega_H l)t+il\varphi}
\phi_{\omega,l}({\bf x})~~~,
\end{equation}
\begin{equation}
\label{1.3}
i\xi\phi_{\omega,l}(t,\varphi,{\bf x})=
\omega\phi_{\omega,l}(t,\varphi,{\bf x})~~~,
\end{equation}
\begin{equation}
\label{1.4}
-i\partial_\varphi\phi_{\omega,l}(t,\varphi,{\bf x})=
l\phi_{\omega,l}(t,\varphi,{\bf x})~~~,
\end{equation}
where ${\bf x}$ are the rest coordinates  of $\cal M$. The
corresponding single-particle excitation of a scalar field has energy
$\omega$ (defined with respect to the Killing vector $\xi$) and the
integer angular momentum $l$.  In the co-rotating coordinates the wave
functions (\ref{1.2}) take the familiar form
\begin{equation}
\label{1.11}
\phi_{\omega,l}(t,\tilde{\varphi}+\Omega_H t,{\bf x})=
e^{-i\omega t+il\tilde{\varphi}}
\phi_{\omega,l}({\bf x})~~~.
\end{equation}

Equation for the spectrum $\omega$ follows from (\ref{1.6}) after
substitution function (\ref{1.2}).
One easily finds the relation
\begin{equation}\label{1.7}
\left[{1 \over N^2}(\omega+l \tilde{N}^\varphi)^2-{1 \over
g_{\varphi\varphi}} l^2-\Delta_{\bf x}-V\right]\phi_{\omega,l}({\bf x})
=0~~~,
\end{equation}
where
\begin{equation}\label{1.8}
\Delta_{\bf x}\equiv-{1 \over \sqrt{-g}}\partial_i\left[
\sqrt{-g}g^{ik}\partial_k\right]~~~.
\end{equation}
The presence of linear in $\omega$ terms in this equation makes it
difficult to use the standard methods for obtaining the density of 
energy levels for this operator. We shall demonstrate now that this
problem can be reduced to the problem in a static space-time.

\bigskip

\noindent
{\bf Proposition:} {\em
The spectrum of single-particle excitations for the wave operator
(\ref{1.5})--(\ref{1.5aa}) in space-time (\ref{1.6})  is uniquely
defined by the spectrum of single-particle excitations for the wave
operator
\begin{equation}\label{1.5a}
\left[ -\tilde{g}^{\mu\nu}(\tilde{\nabla}_\mu-i \lambda A_\mu)
(\tilde{\nabla}_\nu-i\lambda A_\nu)
+V\right] \phi^{(\lambda)}=0~~~
\end{equation}
 on a {\it static} 
background $\tilde{\cal M}$
with the metric
\begin{equation}\label{1.38}
d\tilde{s}^2=\tilde{g}_{\mu\nu}\, dx^{\mu}\, dx^{\nu}=
-Bdt^2+Cd\varphi^2+g_{ik}dx^i dx^k~~~,
\end{equation}
and the abelian gauge field
\begin{equation}\label{1.20}
A= Wd\varphi~~~,
\end{equation}
provided that metric coefficients $B$ and $C$, and the field potential
$W$ are given by relations (\ref{1.15}), (\ref{1.14}), and
(\ref{1.16}), respectively.}

\bigskip

Equation (\ref{1.5a}) contains a real non-negative  parameter $\lambda$
which can be interpreted as the electric charge of the field
$\phi^{(\lambda)}$.  The correspondence between the spectra means that
a single-particle excitation with the energy $\omega$ for wave operator
(\ref{1.5})--(\ref{1.5aa}) in space-time (\ref{1.6})  is uniquely
related to a single-particle excitation for operator  (\ref{1.5a})
taken at $\lambda=\omega$ and having the same energy  $\omega$. For a
static space-time $B=N^2$,  $C=g_{\varphi\varphi}$, $W=0$ and two
problems are equivalent. In general case, the geometry of space
$\tilde{\cal M}$ differs from the geometry of physical space-time $\cal
M$. To emphasize this difference  we call $A$ and $\tilde{\cal M}$  the
fiducial gauge field and the fiducial background, respectively. The
reduction of our problem to the static one on $\tilde{\cal M}$ makes it
possible a considerable simplification in the computations which we use
in a  moment.

\bigskip

We now prove the above proposition. Let us first rewrite Eq.
(\ref{1.7}) in the following equivalent form
\begin{equation}\label{1.13}
\left[\omega^2-B\left(
\Delta_{\bf x}+{1 \over C}(l-\omega W)^2+V\right)
\right]\phi_{\omega,l}({\bf x})=0~~~,
\end{equation}
where $B$, $C$, and $W$ are given by relations
(\ref{1.15})--(\ref{1.16}). Equation (\ref{1.13}) is  the Fourier
transform of the differential equation 
\begin{equation}\label{1.17} 
\left[\omega^2-B\left(
\Delta_{\bf x}-{1 \over C}(\partial_\varphi
-i\omega W)^2+V\right)
\right]\phi_{\omega}(\varphi,{\bf x})=0~~~,
\end{equation}
\begin{equation}\label{1.18}
\phi_{\omega}(\varphi,{\bf x})=\sum_{l}e^{il\varphi}\phi_{\omega,l}({\bf x})~~~.
\end{equation}
Let us introduce the second order differential operator on
a $D-1$-dimensional space 
\begin{equation}\label{1.19}
H^2(\lambda)=B\left(
\Delta_{\bf x}-{1 \over C}(\partial_\varphi
-i\lambda W)^2+V\right)
~~~,
\end{equation}
where $\lambda$ is a real parameter.
Let $\phi^{(\lambda)}_\omega$ be eigen-functions of
$H^2(\lambda)$
\begin{equation}\label{1.21}
H^2(\lambda)\phi^{(\lambda)}_\omega=\omega^2\phi^{(\lambda)}_\omega~~~.
\end{equation}
Obviously, the eigen-functions $\phi^{(\lambda)}_\omega$ enable one to
solve the eigen-problem (\ref{1.17}) because
\begin{equation}\label{1.22}
\phi_\omega(\varphi,{\bf x})=\phi^{(\omega)}_\omega(\varphi,{\bf x})~~~.
\end{equation}
Let us define now the field
\begin{equation}\label{1.22b}
\phi^{(\lambda)}(t,\varphi,{\bf x})=\sum_{\omega} e^{-i\omega t}
\phi^{(\lambda)}_\omega(\varphi,{\bf x})~~~.
\end{equation}
Then it is not difficult to see that the eigen-value problem
(\ref{1.21}) is equivalent to the Klein-Gordon equation (\ref{1.5a}) in
the space-time (\ref{1.38}) with the Abelian field (\ref{1.20}).  To
come to this conclusion one has to use the identity
\begin{equation}\label{1.15a} 
BC=N^2 g_{\varphi\varphi}~~~,  
\end{equation} 
which is the consequence of (\ref{1.14}) and (\ref{1.15}).
By using (\ref{1.21}) and (\ref{1.22}) one also obtains
the density of the energy-levels $\omega$ in (\ref{1.18}),
$dn(\omega) / d\omega$,
as the density of the energy levels
$dn^{(\lambda)}(\omega) /d\omega$ of the operator $H^2(\lambda)$ at
$\lambda=\omega$
\begin{equation}\label{1.23}
{dn(\omega) \over d\omega}=\left. {dn^{(\lambda)}(\omega) \over
d\omega}\right|_{\lambda=\omega}~~~.
\end{equation}
This equation completes the proof of the proposition.

\bigskip

\noindent
Equation (\ref{1.23}) is important because
the density of the energy-levels plays a crucial role in the definition
of the free energy of the system
\begin{equation}\label{1.24}
F[\beta] = \eta\beta^{-1}\int d\omega {dn(\omega) \over d\omega}
\ln (1-\eta e^{-\beta \omega})~~~,
\end{equation}
where $\beta$ is the inverse
temperature and $\eta=+1$ for bosons and $\eta=-1$ for fermions.
Finding the quantity $dn^{(\lambda)}(\omega)/d\omega$
enables one to determine all statistical-mechanical characteristics
of the canonical ensemble on the axially-symmetric background,
including the entropy.

\bigskip

At this point several remarks are in order. As follows from
(\ref{1.19}), the operator $H^2(\lambda)$ is positive when $B= -\xi^2
>0$, that is in the region of the black-hole exterior lying between the
horizon and the null `cylinder', a surface where the co-rotation
velocity reaches the velocity of light.  Outside of this region Eq.
(\ref{1.17}) may not have solutions for real values of energies
$\omega$. This property is the manifestation of the superradiance
phenomenon in the gravitational field of a rotating black hole. In the
presence of superradiance, there does not exist a stationary regular 
in the black hole exterior quantum state. In order to escape the
problem connected with the superradiant modes it is possible to
introduce a mirror-like boundary surrounding the black hole. One can
define a canonical ensemble for the quantum field inside such a
boundary, provided it is chosen to be close enough to the black hole
(inside the null `cylinder'). In what follows we assume that such a
boundary does exist. Note, however, that we shall be interested in the
entropy which is determined by the region in the vicinity of the
horizon, and, hence, the leading divergent contribution to the entropy
does not  depend on the outer boundary.

Also it should be emphasized once again why we define the energy of the
canonical ensemble with respect to the Killing field $\xi$
rather than vector $\partial_t$. The reason is that our final
goal is to compute the entanglement entropy $S^{SM}$ of fields.
The origin of the entanglement entropy is closely related to the
presence of the horizon and the structure of the Killing field $\xi$.
In case of black hole the entanglement entropy 
has a thermal nature and it can be defined as the entropy of 
the thermal atmosphere around a black hole. More formally, it can be 
shown that the entanglement density matrix for a rotating black hole is 
$\hat{\rho}\sim \exp (-\hat{H}/T_H)$ where $\hat{H}$ is the generator 
of canonical transformations along the Killing field $\xi$
and $T_H$ is the Hawking temperature, see, e.g., 
\cite{Israel}.  In this regard, our approach is different from 
the approach \cite{HoKa} where the energy is determined with 
respect to the vector $\partial_t$. In the latter case one always has 
to deal with the contribution of the superradiant modes which appear 
because $\partial_t$ is space-like near the horizon.  
Thus, although the vector $\partial_t$ can be used to define the energy
at spatial infinity, it is not related to the notion of entanglement 
entropy.

\bigskip

As follows from (\ref{1.17}), $H^2(\lambda)$ is Hermitean operator with
respect to the inner product
\begin{equation}\label{1.25}
(\phi_1,\phi_2)=\int d\varphi d^{D-2}{\bf x} \sqrt{-g B^{-2}}~
\phi^{*}_1(\varphi,{\bf x}) \phi_2(\varphi,{\bf x})~~~.
\end{equation}
Following the procedure elaborated in the case
of static space-times \cite{F:98} it is convenient to introduce another
representation of $H^2(\lambda)$
\begin{equation}\label{1.26}
\bar{H}^2(\lambda)=e^{-{D-2 \over 2}\sigma}H^2(\lambda)
e^{{D-2 \over 2}\sigma}~~~,
\end{equation}
\begin{equation}\label{1.27}
e^{-2\sigma}=B~~~.
\end{equation}
In the new representation
\begin{equation}\label{1.28}
\bar{H}^2(\lambda)=-\bar{g}^{ab}
(\bar{\nabla}_a -i\lambda A_a)
(\bar{\nabla}_b -i\lambda A_b)
+\bar{V}~~~,
\end{equation}
$$
\bar{V}=BV+{D-2 \over 2}\left({D-2 \over 2}(\bar{\nabla}\sigma)^2-
\bar{\nabla}^2\sigma
\right)=
$$
\begin{equation}\label{1.29}
B\left[V+{D-2 \over 2}\left(\nabla^\mu w_\mu-{D-2 \over 2}w^\mu w_\mu
\right)\right]~~~.
\end{equation}
The indexes $a,b$ in Eq. (\ref{1.28}) run from 1 to $D-1$
and connections $\bar{\nabla}_a$ are
determined for the metric
\begin{equation}\label{1.30}
d\bar{l}^2=\bar{g}_{ab}dx^a dx^b=
{1 \over B}\left(C d\varphi^2+ g_{ij}dx^i dx^j\right)~~~.
\end{equation}
As earlier the fiducial vector $A_a$ is defined by Eq. (\ref{1.20}).
Finally, the vector $w_\mu$ in (\ref{1.29}) is
\begin{equation}\label{1.30a}
w_\mu=\frac 12 \nabla_\mu\ln B~~~,
\end{equation}
This vector can be interpreted as an acceleration of a static observer
in the fiducial space-time $\tilde{\cal M}$. 

The operator $\bar{H}^2(\lambda)$ is Hermitean with respect
to the standard inner product
\begin{equation}\label{1.31}
(\phi_1,\phi_2)=\int d\varphi d^{D-2}{\bf x} \sqrt{\bar{g}}~
\phi^{*}_1(\varphi,{\bf x}) \phi_2(\varphi,{\bf x})~~~.
\end{equation}

In the static space-time $B=-g_{tt}$ and $w_\mu$ coincides with the
acceleration in the physical space-time. In  four dimensional static
space-time Eq. (\ref{1.35}) reproduces the result of \cite{FF:98a},
\cite{F:98}.

The operator $\bar{H}(\lambda)$ is the Hamiltonian of a relativistic
particle which propagates in the space $\bar{\cal B}$ with the metric
$\bar{g}_{ab}$. The effect of the rotation of the initial space-time is
encoded in the properties of the background metric and in the presence
of an additional gauge field $A_a$.  In the static limit the
operator $\bar{H}(\lambda)$ does not depend on $\lambda$ and coincides
with the single-particle Hamiltonian considered in Refs. \cite{FF:98a},
\cite{F:98}.

\subsection{Spinor fields}

We now show that the reduction of the stationary problem
to the problem on a fiducial static space is universal
and also possible 
for fields with non-zero spins. As an important example we 
consider spinor fields $\psi$ obeying the Dirac equation 
\begin{equation}\label{1.40}
\left(\gamma^\mu \nabla_\mu+m\right)\psi=0~~~.
\end{equation}
The spinor derivatives are $\nabla_\mu=\partial_{\mu}+\Gamma_\mu$,
where $\Gamma_\mu$ are the connections. From now on we work
in the co-rotating frame of reference described by the metric
(\ref{1.10}) and define the connections with respect to this metric.
By choosing in these coordinates the appropriate basis of the
one-forms we can define
the $\gamma$-matrices
\begin{equation}\label{1.41}
\gamma^t={1 \over \sqrt{B}}\bar{\gamma}^t+{\sqrt{C} \over N^2}
\tilde{N}^\varphi\bar{\gamma}^\varphi~~~,~~~
\gamma^\varphi={1 \over \sqrt{C}}\bar{\gamma}^\varphi~~~,
\end{equation}
where $B$ and $C$ are given by (\ref{1.15}) and (\ref{1.14}).
The matrices $\bar{\gamma}^t$ and $\bar{\gamma}^\varphi$ are the
standard Dirac $\gamma$-matrices in the corresponding representation,
\begin{equation}\label{1.42}
(\bar{\gamma}^t)^2=-1~~~,~~~
(\bar{\gamma}^\varphi)^2=1~~~,~~~
\{\bar{\gamma}^\varphi,\bar{\gamma}^t\}=0~~~.
\end{equation}
With this definition one has
\begin{equation}\label{1.43}
(\gamma^t)^2=g^{tt}~~~,~~~
(\gamma^\varphi)^2=g^{\varphi\varphi}~~~,~~~
\{\gamma^t,\gamma^\varphi\}=2g^{t\varphi}~~~,~~~
\{\gamma^i,\gamma^j\}=2g^{ij}~~~,
\end{equation}
where all $\gamma^i$  anticommute with $\gamma^t$ and $\gamma^\varphi$.

We are interested in single-particle excitations 
of the spinor field  which are
the eigen-functions of the Killing vector $\xi$
\begin{equation}\label{1.44}
\psi(\varphi,t,{\bf x})=e^{-i(\omega+\Omega_H l)t}e^{il\varphi}
\psi_{\omega,l}({\bf x})=
e^{-i\omega t}e^{il\tilde{\varphi}}
\psi_{\omega,l}({\bf x})~~~.
\end{equation}
By following the method used for the scalar fields we make
the Fourier transform
\begin{equation}\label{1.46}
\psi_{\omega}(\varphi,{\bf x})=\sum_l e^{il\varphi}
\psi_{\omega,l}({\bf x})~~~.
\end{equation}
The direct computation gives
\begin{equation}
\gamma^t \Gamma_t+\gamma^\varphi \Gamma_\varphi=
\frac 14 \gamma^i\nabla_i\ln (BC)~~~.
\end{equation}
By using this identity it can be shown that the
 equation for $\psi_{\omega}$ obtained from (\ref{1.40}) with the
help of (\ref{1.41}) takes the simple form
\begin{equation}\label{1.45}
\left[(-i\omega){1 \over \sqrt{B}}\bar{\gamma}^t
+\left(\gamma^a(\tilde{\nabla}_a-i\omega A_a-\frac 12
\nabla_a\sigma)+m\right)\right]\psi_{\omega}=0~~~,
\end{equation}
where $a=\{\varphi,i\}$, and $i=1,..,D-2$.
The quantities $A_a$ and $\sigma$ are defined by 
(\ref{1.20}) and (\ref{1.27}), respectively. The spin-connections
$\tilde{\nabla}_a$ are
computed with respect to the metric
\begin{equation}\label{1.47}
dl^2=Cd\varphi^2+g_{ij}dx^idx^j~~~.
\end{equation}
Note that the $i$-th component of $\tilde{\nabla}_a$ coincides
with $i$-th component of the spin-connection $\nabla_\mu$ in the
physical space-time (\ref{1.10}).  The spectral
problem (\ref{1.45}) can be solved by introducing the
fiducial Hamiltonian for spin 1/2 fields
\begin{equation}\label{1.48}
H(\lambda)=i\sqrt{B}\bar{\gamma}^t
\left[\gamma^a(\tilde{\nabla}_a-i\lambda A_a-\frac 12
\nabla_a\sigma)+m\right]~~~.
\end{equation}
The eigen-spinors of $H(\lambda)$,
\begin{equation}\label{1.49}
H(\lambda)\psi^{(\lambda)}_\omega=
\omega\psi^{(\lambda)}_\omega~~~,
\end{equation}
give the eigen-spinors for Eq. (\ref{1.45})
\begin{equation}\label{1.50}
\psi_\omega(\varphi,{\bf x})=\psi^{(\omega)}_\omega(\varphi,{\bf x})~~~.
\end{equation}
In the complete analogy with the case of the scalar fields,
one can define fiducial $D$-dimensional spinors
\begin{equation}\label{1.51}
\psi^{(\lambda)}(t,\varphi,{\bf x})=\sum_\omega e^{-i\omega t}
\psi^{(\lambda)}_\omega(\varphi,{\bf x})~~~
\end{equation}
which obey the Dirac equation
\begin{equation}\label{1.52}
[\tilde{\gamma}^\mu(\tilde{\nabla}_\mu-i \lambda A_\mu)
+m]\psi^{(\lambda)}=0
\end{equation}
on fiducial static space-time $\tilde{\cal M}$ with the
metric (\ref{1.38}) and interact with gauge field (\ref{1.20}).
We see, therefore, that the form of the fiducial background and 
the gauge field is universal for fields of different spins.
This fact may be especially important for supersymmetric models. 

The analysis of spinor fields goes along the lines
of the work \cite{F:98}. Firstly, one can see that the spinor
Hamiltonian (\ref{1.48}) is Hermitean with respect to the
inner product
\begin{equation}\label{1.53}
(\psi_1,\psi_2)=\int d\varphi d^{D-2}{\bf x}\sqrt{-g
B^{-1}}(\psi_1(\varphi,{\bf x}))^{+} \psi_2(\varphi,{\bf x})~~~.
\end{equation}
Secondly, for the further convenience,
one can go to another representation
\begin{equation}\label{1.54}
e^{-{D-1 \over 2}\sigma}\bar{H}(\lambda)
e^{{D-1 \over 2}\sigma}=H(\lambda)~~~,
\end{equation}
\begin{equation}\label{1.55}
\bar{H}(\lambda)=i\bar{\gamma}^t\left[\bar{\gamma}^a
(\bar{\nabla}_a-i\lambda A_a) +m
e^{-\sigma}\right]~~~,
\end{equation}
where the parameter $\sigma$ was introduced in (\ref{1.27}) and
$\gamma$-matrices and spin connections are defined with
respect to metric (\ref{1.30}), which is conformally
related to metric (\ref{1.47}).
The operator $\bar{H}(\lambda)$ is Hermitean with respect to
the standard inner product
\begin{equation}\label{1.56}
(\psi_1,\psi_2)=\int d\varphi d^{D-2}{\bf x}\sqrt{\bar{g}}
(\psi_1(\varphi,{\bf x}))^{+} \psi_2(\varphi,{\bf x})~~~.
\end{equation}
The density of energy-levels $dn^{(\lambda)}(\omega)/d\omega$
of the spinor Hamiltonian
can be computed with the help of relation (\ref{1.34})
by using of the heat kernel of
the operator
\begin{equation}\label{1.57}
\bar{H}^2(\lambda)=-\bar{g}^{\alpha\beta}
(\bar{\nabla}_\alpha -i\lambda A_\alpha)
(\bar{\nabla}_\beta -i\lambda A_\beta)
+\bar{V}(\lambda)~~~,
\end{equation}
\begin{equation}\label{1.58}
\bar{V}(\lambda)=\frac 14 \bar{R}+
B\left(m^2-m{\gamma}^\mu{\nabla}_\mu\sigma
+\frac i2 \lambda \gamma^\mu\gamma^\nu F_{\mu\nu}\right)~~~,
\end{equation}
where $\bar{R}$ is the curvature of the space (\ref{1.30}) and
$F_{\mu\nu}=A_{\nu,\mu}-A_{\mu,\nu}$
is the Maxwell tensor for the fiducial vector potential.

\section{Properties of $\bar{H}^2(\lambda)$
and divergences related to the horizon}
\setcounter{equation}0

From now on we restrict the discussion by the four-dimensional 
space-times ($D=4$). However the analysis can be carried in higher 
dimensions as well.  

Both scalar and spinor single-particle 
Hamiltonians $\bar{H}(\lambda)$ are defined on the space (\ref{1.30}). 
By following the conventions adopted in Ref. \cite{FF:98a} we denote 
this space $\bar{\cal B}$.  In the vicinity of the horizon the geometry 
of $\bar{\cal B}$ is simple.  If $\rho$ is the proper distance to the 
horizon then (see Appendix)
\begin{equation}\label{1.32} 
N^2\simeq \kappa^2 
\rho^2~~~,~~~C\simeq g_{\varphi\varphi}~~~, ~~~B\simeq N^2~~~, 
\end{equation}
where $\kappa$ is the surface gravity of the horizon.
Let $D=4$. By using these asymptotics and Eqs. (\ref{1.29}) and 
(\ref{1.58})  one finds that the potential terms  
at the horizon act as a tachionic mass,
$\bar{V}=-\kappa^2$ for scalars and $\bar{V}=-\frac 32 \kappa^2$
for spinors. 
The presence of the tachionic mass, 
however, is exactly compensated by the mass gap which appears when a 
particle moves on the space $\bar{\cal B}$.  Near the horizon $\rho=0$ 
the metric of $\bar{\cal B}$ takes the form 
\begin{equation}\label{1.33}
d\bar{l}^2\simeq {1 \over \kappa^2\rho^2} (d\rho^2 + d\Omega^2)~~~,
\end{equation}
where in the limit
$\rho\rightarrow 0$ the metric $d\Omega^2$ coincides with the metric
on the horizon.
In this limit the curvature of $\bar{\cal B}$ is constant,
$\bar{R}=-6\kappa^2$ and the space looks as a the hyperbolic
(Lobachevsky) manifold. 
Let us emphasize that these properties
are the same as 
for static space-times \cite{CVZ:95},\cite{BCZ:96}. 

These properties are sufficient to conclude that $\bar{H}^2(\lambda)$
has a continuous non-negative spectrum without a mass gap.
Thus, the density of eigen-values
$dn^{(\lambda)}(\omega) /d\omega$ is divergent  and requires a
regularization. To calculate this
quantity and investigate its divergence we use the method
\cite{F:98} based on the relation
\begin{equation}\label{1.34}
\mbox{Tr}~e^{-\bar{H}^2(\lambda) t}=\int_0^{\infty}
d\omega {d n^{(\lambda)} (\omega) \over d\omega}e^{-\omega^2 t}~~~.
\end{equation}
The density
$dn^{(\lambda)}(\omega) /d\omega$ can be found from (\ref{1.34})
in terms of the trace of the operator $\bar{H}^2(\lambda)$
by using the inverse Laplace transform. The trace involves the
integration over the non-compact space $\bar{\cal B}$.
The volume element of $\bar{\cal B}$
diverges at small $\rho$ as $\rho^{-3}$ and
this is the reason of the divergences of the density of levels.

As was explained in Ref. \cite{F:98},
to study this divergence it is sufficient to restrict oneself only 
by the asymptotic form of the diagonal element of the heat kernel at 
small values of the parameter $t$ 
\begin{equation}\label{1.35} 
\left[ 
e^{-\bar{H}^2(\lambda)t} \right]_{\mbox{diag}} \simeq {1 \over (4\pi 
t)^{3/2}}\left(1+\bar{a}_{1}(\lambda)t+ 
\bar{a}_{2}(\lambda)t^2+...\right)~~~.
\end{equation}
At this point one can make an important observation.
The gauge potential $A_\alpha$ appears in the heat kernel only
in the gauge invariant combinations. Moreover, the coefficient
$\bar{a}_{1}$  does not depend on $A_\alpha$ and it is
is the same as in the case $\lambda=0$. The coefficient
$\bar{a}_{2}(\lambda)$ includes the Maxwell Lagrangian constructed
of $A_\alpha$. The latter term vanishes as $\rho^4$ and it does
not bring the divergence to the trace at small $\rho$.
The same happens in the
higher order coefficients which vanish at least as fast as
$\bar{a}_2(\lambda)$.

Thus, we come to the conclusion that in four-dimensional
space-time the fiducial gauge field does not change the
divergence. If one is interested only in the divergent part
of density of levels, the parameter $\lambda$ in the energy
operator $H^2(\lambda)$
can be put equal to zero.
This fact reduces our problem to Eqs. (\ref{1.5a}), (\ref{1.52})
on the static space-time $\tilde{\cal M}$ with the gauge field  
neglected.  
The divergence of the density of levels can be now computed 
by using the results of \cite{FF:98a}, \cite{F:98} and
expressed in terms of the geometrical characteristics
of $\tilde{\cal M}$ near the horizon.

\bigskip
\noindent
To put it in a more formal way, in four dimensions 
the regularized divergent part of the density of levels
of a field near a rotating black hole
\begin{equation}\label{1.59}
\left[{dn(\omega|\mu) \over d\omega}\right]_{\tiny{div}}=
\left[{dn^{(\lambda=0)} (\omega|\mu)\over 
d\omega}\right]_{\tiny{div}}~~~, 
\end{equation} 
where $\mu$ is a 
regularization parameter. By working, for instance, in the 
Pauli-Villars regularization and by using the expressions of Refs.
\cite{FF:98a}, \cite{F:98}, one finds 
\begin{equation}\label{1.60} 
\left[{dn_s^{(0)} (\omega|\mu)\over d\omega}\right]_{\tiny{div}}
={1 \over (4\pi)^2\kappa}\int_{\Sigma}\left[2b+a
\left({\omega^2 \over \kappa^2}{\cal P}+2\left(\frac 16-\xi\right) R
\right)\right]~~~,
\end{equation}
\begin{equation}\label{1.61}
\left[{dn_d^{(0)} (\omega|\mu)\over d\omega}\right]_{\tiny{div}}
={r_d \over (4\pi)^2\kappa}\int_{\Sigma}\left[2b+a
\left({\omega^2 \over \kappa^2}{\cal P}+{R\over 6}+
{{\cal Q} \over 4}\right)\right]~~~.
\end{equation}
Expressions (\ref{1.60}) and (\ref{1.61}) are referred to the scalar and spinor
densities of levels, respectively, $r_d$ being the dimensionality of the
spinor representation.
The integrals in these expressions
are taken over the bifurcation surface $\Sigma$ of the horizon. As we show in
Appendix, the curvatures of the physical space-time $\cal M$ and the
fiducial one $\tilde{\cal M}$ coincide near $\Sigma$. Thus, the 
quantity $R$ in (\ref{1.60}), (\ref{1.61}) can be identified with the 
scalar curvature of $\cal M$, while other quantities can be written in 
terms of the Riemann and Ricci tensors of $\cal M$ 
\begin{equation}\label{1.62}
{\cal P}=2{\cal R}-{\cal Q}~~~,~~~{\cal Q}=P^{\mu\nu}R_{\mu\nu}~~~,
~~~{\cal R}=P^{\mu\nu}P^{\lambda\rho}R_{\mu\lambda\nu\rho}~~~,
\end{equation}
where $P^{\mu\nu}=l^\mu l^\nu-p^\mu p^\nu$ is a projector onto a two-dimensional
surface orthogonal to $\Sigma$, and $p^\mu$, $l^\mu$ are
two mutually orthogonal normals of $\Sigma$ ($l^2=-p^2=1$).
The regularization parameter $\mu$ defines the scale of the Pauli-Villars
masses, and at large $\mu$
\begin{equation}\label{1.63}
a\simeq \ln{\mu^2 \over m^2}~~~,~~~
b\simeq \mu^2\ln {729 \over 256} -m^2\ln {\mu^2 \over m^2}
\end{equation}
where $m$ is the mass of the field (see for details \cite{FF:98a}).

Note that the form of these equations is completely the same as in the static
space-times. By using (\ref{1.60}), (\ref{1.61}) 
one can find the divergences of the free energy
of the fields and the entropy which just repeat expressions (4.26) and
(4.27) of Ref. \cite{FF:98a}. For instance, in the Pauli-Villars
regularization the divergence of the
entropy of the quanta near the horizon
is given by the expression
\begin{equation}\label{1.64}
S_{\tiny{\mbox{div}}}={\eta \over 48 \pi}
\int_{\Sigma}[bf_1+a(2p_1 {\cal P}+p_2 R+p_3{\cal Q})]~~~.
\end{equation}
For scalars
$\eta=1$, $f_1=1$, $p_1=1/60$, $p_2=1/6-\xi$, $p_3=0$; for spinors
$\eta=-1$ , $f_1=-r_d/2$, $p_1=-7r_d/480$, $p_2=r_d/24$, $p_3=-r_d/16$.
The entropy is evaluated at the Hawking temperature
$\beta_H^{-1}={\kappa/ 2\pi}$. For scalar fields the same result was
recently obtained  by the WKB method in \cite{JiYa}-\cite{JiYa:a}. Also
in the scalar case one can  find the divergent part of the entropy by
using the Euclidean formalism (conical singularity method), see
\cite{MaSo:96}.

Analogous results can be found in the dimensional regularization. It
should be noted that the divergences caused by the presence of the
horizon can be also regularized by using the infrared type cutoff. In
this regularization one just cuts all integrations near the horizon at
some proper distance, see for a review \cite{FF:98a}. Our results can
be used to find explicit expressions for the entropy in this case,
however, the discussion of this regularization is beyond the scope of
this paper. For rotating black hole space-times this question was
studied in Refs. \cite{LeKi}--\cite{HoKa}.

\bigskip

The fact that divergent part (\ref{1.64}) of the
entropy of quantum  fields near a rotating 
black hole has the same form as for a static black hole has a number
of immediate consequences. One of the consequences is that
for minimally coupled fields divergence (\ref{1.64}) is 
completely removed
by the standard renormalization of the gravitational couplings
(including the Newton constant) in the bare tree-level part
of the black hole entropy.  The proof of this statement for 
static black holes can be found in Refs. \cite{SU}--\cite{DLM:95}
and it is generalized without changes to rotating black holes.
Another application of our results is the problem
of black hole entropy in models of induced gravity.

\section{Rotating black holes in induced gravity}
\setcounter{equation}0

The models of induced gravity \cite{FFZ:97}--\cite{FF:97} 
were constructed with the purpose to understand the mechanism of the
generation the Bekenstein-Hawking entropy of black holes
in the situation when the low-energy gravity is induced by quantum 
effects. It was argued that for a Schwarzschild black hole
the ultraheavy fields (constituents) which induce the
Einstein gravity in the low-energy limit are  
microscopic degrees of freedom which are responsible for the
Bekenstein-Hawking entropy $S^{BH}$. The important
requirement of models \cite{FFZ:97}--\cite{FF:97} is the
absence of the leading ultraviolet divergences, which imposes
constraints on the parameters of the constituents.
By using these constraints one finds the relation
between $S^{BH}$ and the entropy $S$ of 
the constituents propagating near the black hole horizon  
\begin{equation}\label{3.1}
S^{BH}=S-Q~~~.
\end{equation}
The quantity $Q$ is the quantum average of the Noether charge
\cite{FF:97},\cite{F:98b} which appears because of non-minimal
couplings of the constituents with the curvature. Such couplings are 
necessary to provide  cancellation of the leading ultraviolet
divergences in the  induced effective action. It is important that the
same couplings provide finiteness of the induced Bekenstein-Hawking
entropy (\ref{3.1}): the divergence (\ref{1.64}) of the entropy $S$ of
the constituents is compensated by the divergence of the Noether charge
$Q$.

We now have all means to generalize result (\ref{3.1}) of 
\cite{FFZ:97}-\cite{FF:97} to Kerr black holes. Consider  induced
gravity models with spinor and  non-minimally coupled scalar
constituents only.  The constraints on the parameters of the
constituents and proof of relation (\ref{3.1}) for a Schwarzschild
black hole  are given in \cite{FFZ:97}.  The Kerr black hole is the
vacuum solution and the geometrical structure of the divergences in the
effective action for the Kerr and Schwarzschild backgrounds are
identical. The induced effective action  for a Schwarzschild solution
contains logarithmic divergences of a  topological form only. These
divergences play no role and can be  neglected \cite{FFZ:97}.  We
conclude that the same property is true  for the action on the Kerr
background. In this sense the induced  gravity
\cite{FFZ:97}--\cite{FF:97} for vacuum static and rotating  black hole
is ultraviolet finite theory.

Consider now the divergence of the entropy $S$ for a scalar
or spinor constituent, see Eq. (\ref{1.64}). According to
Eqs. (\ref{2.13}), (\ref{2.21}) of Appendix A,
\begin{equation}\label{3.2}
{\cal Q}=0~~~,~~~\int_{\Sigma} {\cal R}=8\pi
\end{equation} 
for the Kerr background. Thus, 
\begin{equation}\label{3.3}
S_{\tiny{\mbox{div}}}={\eta \over 48 \pi}
bf_1{\cal A}+C~~~,
\end{equation}
where ${\cal A}=\int_{\Sigma}$ is the area of the black hole horizon
and $C$ is a divergent numerical constant
(which is not observable and can be neglected). Thus, the entropy of 
the constituents in the leading order 
is proportional to the area of the horizon of the Kerr black hole
and looks similar to the Bekenstein-Hawking entropy.
Equation (\ref{3.3}) has precisely the same form as the entropy 
for a Schwarzschild black hole. 
As far the Noether charge $Q$ is concerned, in the considered models it 
is determined by the averages of the scalar operators $\langle 
\hat{\phi}^2\rangle$ on the horizon $\Sigma$. 
In quantum states where the Green functions
are analytical on the horizon 
\begin{equation}\label{3.4}
\int_{\Sigma}\langle \hat{\phi}^2 \rangle={1 \over 16 \pi^2}
b{\cal A}~~~,
\end{equation}
where the function $b$ is given in Pauli-Villars 
regularization by (\ref{1.63}).
This equation holds on all vacuum backgrounds in the leading
order approximation, and one can conclude that the Noether  
charges $Q$ for the two black holes have the same form.

These observations show that in induced gravity models
Eq. (\ref{3.1}) does hold for the Bekenstein-Hawking entropy 
of a Kerr black hole. By using (\ref{1.63}), (\ref{3.3}), (\ref{3.4}) 
in (\ref{3.1}) one can check how 
divergence of $Q$ compensates the divergence of $S$ and one gets a 
finite expression which coincides with the induced entropy $S^{BH}$.  
It is a strong support of universality of the 
statistical-mechanical explanation of the Bekenstein-Hawking entropy in 
induced gravity.

\bigskip

We complete this Section with remarks concerning the interpretation
of the Noether charge $Q$. The origin of subtraction
in (\ref{3.1}) can be explained as follows \cite{FF:97}.
The Bekenstein-Hawking entropy of a rotating black hole in induced 
gravity is related to 
the spectrum of the black hole mass $M$ and angular momentum $J$
which determine the grand canonical ensemble.
On the other hand the 
statistical-mechanical entropy $S$ is determined by the 
spectrum of the Hamiltonian ${\cal H}_\xi$ of the constituents. 
Operator ${\cal H}_\xi$ is the generator of canonical transformations 
of the system along the Killing field $\xi$. In the 
presence of non-minimal couplings these spectra are different  
and subtraction of $Q$ in (\ref{3.1}) is required to go
from one spectrum to another.

To make this statement more clear consider a 
small excitation of constituent field having energy $\cal E$
and angular momentum $\cal J$ over a vacuum with ${\cal E}
={\cal J}=0$. Such an excitation 
results in a change of the black 
hole mass $M$ and angular momentum $J$
\begin{equation}\label{3.5}
\delta M=T_H\delta S^{BH}+\Omega_H \delta J
+{\cal E}-\Omega_H {\cal J}~~~,
\end{equation}
see \cite{F:98b}.
Strictly speaking, this relation implies definition of 
$M$ and $J$ at spatial infinity.
For this reason,
the energy and angular momentum of fields are the integrals
over all black hole exterior \cite{F:98b}
\begin{equation}\label{3.6}
{\cal E}=\int_{\Sigma_t} T^{\mu\nu}t_\mu d\Sigma_\nu~~~,
\end{equation}
\begin{equation}\label{3.7}
{\cal J}=-\int_{\Sigma_t} T^{\mu\nu}\varphi_\mu d\Sigma_\nu~~~,
\end{equation}
where $T^{\mu\nu}$ is the stress-energy tensor of the fields.
$\Sigma_t$ is the hypersurface of constant time $t$ and
$d \Sigma_t$ is the future-directed vector of the volume element
of $\Sigma_t$.
The components $t_\mu$, $\varphi_\mu$ correspond to
the Killing vector fields $\partial_t$ and $\partial_\varphi$,
respectively.

In the induced gravity approach the constituents which contribute
to the black hole entropy are assumed to be very heavy and have
the mass of the order of the Planckian mass. Since  Hawking and
the superradiant emissions of such particles are exponentially 
suppressed, they are practically trapped inside the potential barrier. 
The latter in many aspects plays the role of the external boundary 
which is required to define the canonical ensemble. 

Thus,
the dominant contribution to integrals (\ref{3.6}), (\ref{3.7}) 
comes from the region inside the null 'cylinder' (see discussion
in Section 2.1) where one can define the energy of constituents
associated to the Killing field 
$\xi=\partial_t+\Omega_H\partial_\varphi$ 
\begin{equation}\label{3.8} 
{\cal E}_\xi={\cal E}-\Omega_H{\cal J}=
\int_{\Sigma_t} T^{\mu\nu}\xi_\mu d\Sigma_\nu~~~.
\end{equation}
After that variational formula (\ref{3.5}) is represented as
\begin{equation}\label{3.9}
\delta M-\Omega_H\delta J=T_H\delta S^{BH}
+{\cal E}_\xi~~~,
\end{equation}
and it looks somewhat similar to the formula for static black holes.
Thus, for a black hole with the fixed area the spectrum of $M$ and $J$
is related to the spectrum of energies ${\cal E}_\xi$
of the constituents near the horizon. The crucial observation is
that the energy ${\cal E}_\xi$ and the Hamiltonian ${\cal H}_\xi$ 
of the
non-minimally coupled constituents differ by a total derivative
which picks up a non-vanishing contribution on the 
inner boundary of $\Sigma_t$, i.e., on the horizon.
The boundary term is 
the Noether charge on $\Sigma$
\begin{equation}\label{3.10}
{\cal H}_\xi-{\cal E}_\xi=T_H Q~~~,
\end{equation}
where $T_H$ is the Hawking temperature.
It is because of Eqs. (\ref{3.9}), (\ref{3.10}) we expect 
that the two entropies, $S^{BH}$ and $S$, are different
and related by (\ref{3.1}). Studying further aspects of the 
subtraction in (\ref{3.1}) repeats the analysis of a 
Schwarzschild black hole and we advise
corresponding work \cite{FF:97} for the interested reader.

\section{Concluding remarks}
\setcounter{equation}0

Our results can be summarized as follows.  We developed a formalism of
statistical-mechanical computations for a canonical ensemble of fields 
near the horizon of a rotating black hole. Such a canonical ensemble
can be defined when the reference frame co-rotates with the angular
velocity of the black hole. We suggested a method how to reduce
computations on the stationary background to computations on a fiducial
static space-time in the presence of a fiducial gauge potential. Our
method enables one to use the known results for this problem and to 
get a number of rigorous results for rotating black holes. We believe
that the method may be helpful in a number of applications, some of
which were discussed in Sections 3 and 4. In particular, it is worth
pointing out here the proof of universality of statistical-mechanical
origin of the  Bekenstein-Hawking entropy of vacuum black holes in the 
models of induced gravity.

\bigskip 

One of the results of our analysis is that the  Euclidean formulation
of the theory based on the conical-singularity method \cite{MaSo:96}
reproduces correctly the divergence of the entropy (\ref{1.64}) for
stationary space-times. In spite of this fact, the equivalence between
the canonical formulation of statistical mechanics and the Euclidean
one remains unclear in this case. Unfortunately, one cannot apply the 
analysis of \cite{F:98} given for static geometries. The difficulty is
related not to the horizon but to the prescription used for the
Euclidean theory which implies  an analytical continuation of some
parameters of the metric. This issue is an interesting problem for
further research.

\vspace{12pt}
{\bf Acknowledgments}:\ \ 
The work of V.F. is  partially supported by the Natural Sciences and
Engineering Research Council of Canada, and  D.F. is supported in part
by the RFBR grant N 99-02-18146.

\newpage

\appendix
\section{Geometry of $\cal M$ and $\tilde{\cal M}$ near the horizon}
\setcounter{equation}0

In this appendix, we consider four-dimensional space-times.
Generalization of the results to higher dimensions is straightforward.
Consider first a line element ($i,j=2,3$)
\begin{equation}\label{2.1a}
dL^2=g_{ij}\, dx^i\, dx^j\, ,
\end{equation}
which enters metrics (\ref{1.10})and (\ref{1.38}) of spaces $\cal M$ and
$\tilde{\cal M}$, respectively, 
\begin{equation}\label{2.1}
ds^2=-Bdt^2+2g_{\varphi\varphi} \tilde{N}_\varphi dt d\tilde{\varphi}
+g_{\varphi\varphi}d\tilde{\varphi}^2
+dL^2~~~,
\end{equation}
\begin{equation}\label{2.2}
d\tilde{s}^2=-Bdt^2+C d\tilde{\varphi}^2 +dL^2~~~.
\end{equation}
Starting with an
arbitrary surface ${\cal S}$ and introducing geodesic 
 coordinates one can always rewrite (\ref{2.1a}) in the form
\begin{equation}\label{2.1b}
dL^2=d\rho^2 +v(\rho,x)\ dx^2\, .
\end{equation}
It is convenient to choose the surface ${\cal S}$ where $\rho=0$ to
coincide with the horizon. For  given metric (\ref{2.1a}) it is
sometimes difficult to solve the geodesic equations required for the
coordinate transformation which results in (\ref{2.1b}). Much easier
problem is the reduction of the metric (\ref{2.1a}) to the form
\begin{equation}\label{2.1c}
dL^2=d\rho^2 +2\, q(\rho,x)\,d\rho\, dx+ v(\rho,x)\ dx^2\, .
\end{equation}
We shall use this form for further calculations.

It can be shown that for a geometry which is regular at the horizon
the following decompositions of the metric coefficients is valid
 near the horizon (i.e., at small $\rho$) 
\begin{equation}\label{2.3}
B=\kappa^2\rho^2(1+b(x)\rho^2+O(\rho^4))~~~,
\end{equation}
\begin{equation}\label{2.4}
g_{\varphi\varphi}\tilde{N}^\varphi=\rho^2 p_1(x)+O(\rho^4)~~~,
\end{equation}
\begin{equation}\label{2.5}
g_{\varphi\varphi}=f_1(x)+f_2(x) \rho^2+O(\rho^4)~~~,
\end{equation}
\begin{equation}\label{2.6}
v=v_1(x)+v_2(x)\rho^2+O(\rho^4)~~~,
\end{equation}
\begin{equation}\label{2.7}
q=\rho q_1(x)+O(\rho^3)~~~,
\end{equation}
\begin{equation}\label{2.8}
C=f_1(x)+\left(f_2(x)+{p_1^2(x) \over \kappa^2}\right) \rho^2+O(\rho^4)
\end{equation}
The constant $\kappa$ is the surface gravity of the black hole horizon.
Equation (\ref{2.8}) follows from Eqs. (\ref{1.9.c}) and (\ref{1.14}).

Now, by direct computation, one can express the components of the Riemann
and Ricci tensors on the horizon in terms of the coefficients present
in (\ref{2.3})--(\ref{2.8}). Let us define on the bifurcation
surface $\Sigma$ of the horizon the
following quantities\footnote{These definitions coincide with 
(\ref{1.62}).} 
\begin{equation}\label{2.9} 
{\cal Q}=P^{\mu\nu}R_{\mu\nu}~~~, ~
~~{\cal R}=P^{\mu\nu}P^{\lambda\rho}R_{\mu\lambda\nu\rho}~~~, 
~~~P^{\mu\nu}=l^\mu l^\nu-p^\mu p^\nu~~~,
\end{equation}
where $P^{\mu\nu}$ is the projector onto two-dimensional surface
orthogonal to $\Sigma$.
For the space $\cal M$,
\begin{equation}\label{2.10}
{\cal R}=2R^t_{~~\rho t \rho}=-6b-2{q_1^2 \over v_1}~~~,
\end{equation}
\begin{equation}\label{2.11}
{\cal Q}=2\left[-3b-{q_1^2 \over v_1}+\left(q_1 \over v_1\right)'-
{v_2 \over v_1}-{1 \over f_1}\left(f_2+{p_1^2 \over \kappa^2}\right)
+{g_1 \over 2v_1}\left({v_1' \over v_1}+{f_1' \over f_1}\right)\right]~~~,
\end{equation}
where $f'\equiv df/dx$.
Then, by using (\ref{2.3}), (\ref{2.6})--(\ref{2.8}) one verifies that for the space
$\tilde{\cal M}$
\begin{equation}\label{2.12}
\tilde{\cal R}={\cal R}~~~,~~~\tilde{\cal Q}={\cal Q}~~~.
\end{equation}
According to the Gauss-Codacci equations
\begin{equation}\label{2.13}
R=R_\Sigma+2{\cal Q}-{\cal R}~~~.
\end{equation}
where $R$ and $R_\Sigma$ are scalar curvatures of $\cal M$ and
$\Sigma$, respectively. (Here we took into account that the extrinsic
curvatures of $\Sigma$ vanish due to the isometry). The same equation
is valid for the scalar curvature $\tilde{R}$ of $\tilde{\cal M}$ and
one concludes that
\begin{equation}\label{2.14}
\tilde{R}=R~~~.
\end{equation}
Therefore, all the curvatures which characterize the geometry of the
physical $\cal M$ and fiducial $\tilde{\cal M}$ space-times coincide at
the horizon $\Sigma$. As far as other geometrical properties (e.g.,
derivatives  of the curvatures at $\Sigma$)  are concerned, they can be
different in general. This fact, however, is not important when one
studies the divergences of the density of energy-levels in
four-dimensional  theory, Section 1.3.

\bigskip

For the sake of completeness, we give the expressions for the surface
invariants for ${\cal Q}$ and ${\cal R}$ for the
Kerr-Newman black hole of  mass $M$, charge $Q$, and 
angular momentum $J=aM$. The metric in Boyer-Lindquist coordinates is
$$
ds^2=-\left(1-{2Mr-Q^2 \over \Sigma}\right)dt^2-
2{(2Mr-Q^2)a\sin^2\theta \over \Sigma}dtd\varphi
$$
\begin{equation}\label{2.15}
+{\Sigma \over \Delta} dr^2+\Sigma d\theta^2+{A \sin^2\theta
\over \Sigma}d\varphi^2~~~,
\end{equation}
\begin{equation}\label{2.16}
\Delta=r^2-2Mr+a^2+Q^2~~~,~~~
\Sigma=r^2+a^2\cos^2\theta~~~,
\end{equation}
\begin{equation}\label{2.17}
A=(r^2+a^2)^2-\Delta a^2\sin^2\theta~~~.
\end{equation}
The horizon is defined by the equation $\Delta=0$ and
is located at
\begin{equation}\label{2.18}
r=r_+=M+\sqrt{M^2-Q^2-a^2}~~~.
\end{equation}
The surface gravity $\kappa$ and the angular velocity
$\Omega_H$ for the Kerr-Newman black hole are
\begin{equation}\label{2.19}
\kappa={r_+-M \over r_+^2+a^2}~~~,~~~\Omega_H={a \over r_+^2+a^2}~~~.
\end{equation}
The Kerr-Newmann metric (\ref{2.15}) can be brought to form
(\ref{2.1}) when one goes to the corotating coordinate frame by the
substitution 
$\varphi=\tilde{\varphi}+\Omega_Ht$. The  coordinates $\theta$
and $x$ in (\ref{2.1}) and (\ref{2.15}) coincide, the coordinate
$\rho$ is determined as
\begin{equation}\label{2.20}
\rho=\int_{r_+}^r dr \,\sqrt{g_{rr}(r,\theta)} ~~~.
\end{equation}
By using Eqs. (\ref{2.15})--(\ref{2.19})
one can find the coefficients $b$, $p_1$, $v_i$, $f_i$, and $q_i$.
After some simple algebra one finds from
Eqs. (\ref{2.10}) and (\ref{2.11})
\begin{equation}\label{2.21}
{\cal Q}=-{2 Q^2 \over \Sigma_+^2}~~~,
\end{equation}
\begin{equation}\label{2.22}
{\cal R}={2 \over \Sigma_+^3}\left(4r_+^2(2Mr_+-Q^2)+
\Sigma_+(Q^2-6Mr_+)\right)~~~,
\end{equation}
where $\Sigma_+$ is the value of $\Sigma$ at $r=r_+$.
The scalar curvature $R$ of the Kerr-Newmann solution
vanishes everywhere.

\end{document}